\documentclass[11pt]{article}
\usepackage{fullpage}
\usepackage[utf8]{inputenc}
\usepackage[english]{babel}
\usepackage[T1]{fontenc}
\usepackage{amsmath,amssymb,amsfonts,amsthm}
\usepackage{url}
\usepackage{hyperref}
\usepackage{cleveref}
\usepackage{graphicx}
\usepackage[ruled,vlined, linesnumbered]{algorithm2e}

\newtheorem*{theorem*}{Theorem}

\usepackage{authblk}

\newcommand{\faults}{{\cal F}}
\newcommand{\checks}{{\cal C}}
\newcommand{\faultset}{\varphi}
\DeclareMathOperator{\MWPM}{MWPM}
\DeclareMathOperator{\UF}{UF}
\newcommand{\Prob}{\mathbb{P}}

\usepackage{xcolor}

\title{Splitting decoders for correcting hypergraph faults}

\author[1]{Nicolas Delfosse}
\author[1]{Adam Paetznick}
\author[1, 2]{Jeongwan Haah}
\author[1, 2]{Matthew B. Hastings}

\affil[1]{Microsoft Quantum, Redmond, Washington, USA}
\affil[2]{Microsoft Station Q, Santa Barbara, California, USA}

\begin{document}

\maketitle

\begin{abstract}
The surface code is one of the most popular quantum error correction codes. It comes with efficient decoders, such as the Minimum Weight Perfect Matching (MWPM) decoder and the Union-Find (UF) decoder, allowing for fast quantum error correction.
For a general linear code or stabilizer code, the decoding problem is NP-hard.
What makes it tractable for the surface code is the special structure of faults and checks: Each X and Z fault triggers at most two checks.
As a result, faults can be interpreted as edges in a graph whose vertices are the checks, and the decoding problem can be solved using standard graph algorithms such as Edmonds' minimum-weight perfect matching algorithm.
For general codes, this decoding graph is replaced by a hypergraph making the decoding problem more challenging.
In this work, we propose two heuristic algorithms for splitting the hyperedges of a decoding hypergraph into edges.
After splitting, hypergraph faults can be decoded using any surface code decoder.
Due to the complexity of the decoding problem, we do not expect this strategy to achieve a good error correction performance for a general code.
However, we empirically show that this strategy leads to a good performance for some classes of LDPC codes because they are defined by low weight checks.
We apply this splitting decoder to Floquet codes for which some faults trigger up to four checks and verify numerically that this decoder achieves the maximum code distance for two instances of Floquet codes.
\end{abstract}

\section{Introduction}

The decoder is an essential building block of a fault-tolerant quantum computer.
Its role is to identify faults occurring during a quantum computation so that they can be corrected before they spread to the whole system.
To avoid this proliferation of errors, the decoder must be fast. 
This significantly restricts the type of quantum error correction codes we can consider for fault-tolerant quantum computing because the decoding problem is generally non-trivial. Finding a most likely error is NP-hard like in case of classical linear codes~\cite{berlekamp1978inherent} and maximum likelihood decoding with stabilizer codes is $\#$P-hard~\cite{iyer2015hardness}.

One of the main reasons for the success of the surface code~\cite{dennis2002topological, raussendorf2007fault, fowler2012surface} is that the corresponding decoding problem is easy: it can be reduced to a matching problem in a graph which can be solved in polynomial time using a standard minimum-weight perfect matching algorithm~\cite{dennis2002topological}.

The main drawback of the surface code is that its encoding rate is vanishing and therefore it leads to a large qubit overhead.
Quantum LDPC codes are promising candidates to reduce the qubit count of large-scale quantum applications because they achieve better parameters than topological codes~\cite{tillich2013quantum, kovalev2012improved, hastings2021fiber, breuckmann2021quantum, panteleev2022asymptotically, leverrier2022quantum}.
Moreover, circuit-level simulations show that one could hope for significant reduction in the number of qubits for a fault-tolerant quantum memory~\cite{tremblay2022constant, higgott2023constructions}.
However, their decoding problem corresponds to a hypergraph matching problem that is more challenging than the corresponding graph problem.
More work is needed to improve their decoders.
The recently discovered good quantum LDPC codes~\cite{panteleev2022asymptotically, leverrier2022quantum} have a linear time decoder~\cite{gu2023efficient, lin2022good, leverrier2023efficient} but explicit code constructions are missing for these schemes.
Classical Belief Propagation (BP) decoders~\cite{mackay2003information, richardson2008modern} do not perform well in general because the Tanner graph of quantum LDPC codes contains many short cycles.
Different strategies have been considered for quantum LDPC codes, either by modifying BP ~\cite{poulin2008iterative, panteleev2021degenerate, roffe2020decoding, grospellier2021combining, du2022stabilizer}, or by adapting the UF decoder~\cite{delfosse2022toward}.
This generally leads to decoders with increased complexity, degraded performance, or both.

Here, we take a different approach. Our goal is not to design a decoder for all quantum LDPC codes. Instead, we start from a matching decoder and aim to make it more flexible in order to extend the range of applicability.
We propose two heuristics that let us apply matching decoders such as the MWPM decoder~\cite{dennis2002topological} or the UF decoder~\cite{delfosse2021almost}, originally designed for surface codes, to cousins of the surface codes such as Floquet surface codes~\cite{hastings2021dynamically, hastings2021dynamically, gidney2022benchmarking, paetznick2022performance}.

\begin{figure}[!ht]
\centering
\includegraphics[scale=.5]{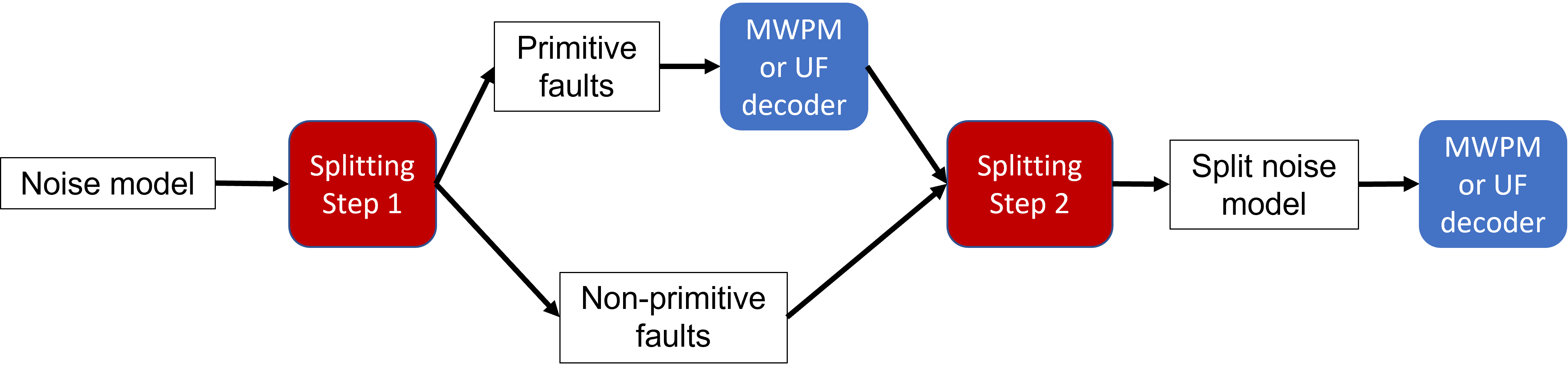}
\caption{Procedure to build a MWPM decoder or a UF decoder for faults triggering more than two checks.
This figure represents the decoder-based splitting method.}
\label{fig:splitting_diagram}
\end{figure}

Our first heuristic is a decoder-based splitting illustrated in Figure~\ref{fig:splitting_diagram}.
First, a set of faults forming a graph is selected. 
It is a subset of the set of all possible faults of the noise model that we call primitive faults.
Because the primitive faults define a graph, one can build a MWPM decoder or a UF decoder for these faults. The non-primitive faults are then split into paths of primitive faults using this decoder.
Our second heuristic is a recursive splitting. We go over the non-primitive faults and remove their primitive parts until it remain only a fault that trigger at most two checks. This fault is then added to the primitive set.

We checked numerically that our (decoder-based) splitting decoder reaches the maximum achievable distance for surface codes and for examples of Floquet surface codes 
(this decoder was a key ingredient in our simulation of Floquet surface codes~\cite{paetznick2022performance}).

In Section~\ref{sec:MWPM}, we review the standard MWPM decoder and explain that the MWPM decoder can be applied to a set of faults such that each fault triggers at most two checks.
In Section~\ref{sec:splitting}, we describe two methods to split faults that trigger more than two checks.
Using this splitting as a preprocessing step, we can build a MWPM decoder and a UF decoder for Floquet codes.

\section{Standard MWPM decoder}
\label{sec:MWPM}

In this section, we review the standard MWPM decoder~\cite{dennis2002topological} and provide a simple description of the algorithm.
This algorithm was extensively optimized over the past two decades improving the time complexity~\cite{fowler2012towards, higgott2023sparse}.

\subsection{Faults and checks}
\label{subsec:faults}

Assume that we are given a system equipped with a set of checks whose role is to detect faults.
In the absence of faults, all the checks return a trivial outcome.
For simplicity, we assume that each check returns a single outcome bit.
To detect and correct faults, we measure the checks and use the set of triggered checks (the checks returning a non-trivial outcome) to identify the faults which occur\footnote{We use the term check, common in classical coding theory, although some authors refer to these as detectors~\cite{gidney2021stim}.}
For a given quantum circuit, one can efficiently generate a set of checks using the algorithms described in~\cite{delfosse2023spacetime}.

\medskip
In what follows, $\checks$ denotes the finite set of checks of the system.
A {\em fault} is an unwanted modification of the system.
We consider a {\em noise model} given by a finite set of independent faults $\faults = \{f_1, \dots, f_m\}$ where each fault occurs with probability $\Prob_{\faults}(f_i)$.
By a {\em fault configuration}, we mean a subset $\faultset \subset \faults$ of faults.
We denote a fault configuration as a formal sum with binary coefficients
$$
\faultset = \sum_{f \in \faults} \varphi_f f
$$
where $\varphi_f = 1$ if $f \in \faultset$ and $\varphi_f = 0$ otherwise.
The sum of two fault sets $\faultset + \faultset'$ where 
$\faultset = \sum_{f \in \faults} \varphi_f f$ and $\faultset' = \sum_{f \in \faults} \varphi_f' f$,
is defined to be the fault configuration
$$
\sum_{f \in \faults} (\varphi_f + \varphi_f') f
$$
where $\varphi_f + \varphi_f'$ refers to the addition modulo 2.
The sum of two fault configurations corresponds to the symmetric difference of the corresponding fault sets.
We use binary coefficient in this formal sum because Pauli faults satisfy $f^2 = I$ and therefore a fault which appears twice cancels out.
We could consider more general noise models by adjusting the coefficient space.

\medskip
Any fault configuration $\faultset$ triggers a set of checks, denoted $\sigma(\faultset) \subset \checks$, that we call the {\em syndrome} of $\faultset$.
Like fault configurations, a syndrome is represented as formal sum of checks and the addition of syndromes is defined similarly.

\medskip
We assume that all the faults $f_i$ have distinct syndromes. If two faults $f_i$ and $f_j$ have the same syndrome, we can remove $f_j$ from $\faults$ and replace $\Prob_{\faults}(f_i)$ by $\Prob_{\faults}(f_i) + \Prob_{\faults}(f_j) - \Prob_{\faults}(f_i)\Prob_{\faults}(f_j)$.
It may happen that $f_i$ and $f_j$ have the same syndrome but have a different action on the system. In this case, the set of checks is not good enough to distinguish $f_i$ and $f_j$. If we care about the difference between these two actions on the system, we should design a different set of checks. 
Similarly, we assume that all faults $f_i$ trigger at least one check.
The faults which do not satisfy this assumption are undetectable and uncorrectable with this set of checks.

\subsection{MWPM decoder for graph-like noise models}

Let us review the MWPM decoder (Algorithm~\ref{algo:mwpm_decoder}).
We consider a noise model satisfying the two following assumptions.
\begin{enumerate}
\item \label{assumption:edges_like_fault}
{Edge-like faults:} Each fault $f_i$ triggers at most two checks.
\item \label{assumption:linearity}
{Check linearity:} For all $\faultset, \faultset' \subset \faults$, we have 
$
\sigma(\faultset + \faultset') = \sigma(\faultset) + \sigma(\faultset').
$
\end{enumerate}
A noise model $\faults$ that satisfies these assumptions is said to be a {\em graph-like noise model}.
The linearity holds for all classical linear codes and for all stabilizer codes. 
More generally, it holds for quantum circuit faults corrected using the checks of the outcome code or the spacetime code as in~\cite{delfosse2023spacetime}. This formalism includes subsystem codes and Floquet codes.
In what follows, we only consider linear checks.
We need only to test that the first assumption is satisfied.

\begin{algorithm}[h!]
\DontPrintSemicolon
\SetKwInOut{Input}{input}\SetKwInOut{Output}{output}
\Input{A syndrome $\sigma \subset \checks$. The decoding graph $G_{\faults}$.}
\Output{A most likely fault configuration $\faultset$ with syndrome $\sigma$.}

\BlankLine
	Initialize $\bar \sigma = \sigma$. \;
	\For {each connected component $C$ of the decoding graph}
	{
		If $C$ contains an odd number of vertices of $\sigma$, add the boundary vertex of the component to $\bar \sigma$. \;
	}
	Construct the distance graph $K_{\bar \sigma}$. \;
	Compute a minimum weight perfect matching $M$ in $K_{\bar \sigma}$. \;
	Initialize a trivial fault configuration $\faultset = 0$. \;
	\For {each edge $\{u, v\} \in M$}
	{
		Compute a set of edges $e_{i_1}, \dots, e_{i_s}$ forming a shortest path from $u$ to $v$ in $G_{\faults}$. \;
		Replace $\faultset$ by $\faultset + f_{i_1} + \dots + f_{i_s}$. \;
	}
	Return $\faultset$.
	\caption{MWPM decoder.}
	\label{algo:mwpm_decoder}
\end{algorithm}

\medskip
The {\em decoding graph} of the noise model $\faults$ is constructed in two steps.
First, we build a graph whose vertex set is the set of checks.
Two checks are connected by an edge if there exists a fault $f_i$ that triggers these two checks.
For each connected component of this graph, we add an extra vertex that we refer to as the {\em boundary vertex} of the component.
Then, for each fault $f_i$ that triggers a single check $c$, we add an edge connecting $c$ with the boundary vertex of its connected component.
By construction, there is a one-to-one correspondence between the faults $f_i$ of $\faults$ and the edges of the decoding graph. The edge associated with $f_i$ is denoted $e_i$.
The decoding graph is a weighted graph and we define the weight $w_i$ of $e_i$ to be 
$$
w_i = -\log \left( \frac{\Prob_{\faults}(f_i)}{1 - \Prob_{\faults}(f_i)} \right) \cdot
$$
The decoding graph associated with $\faults$ is denoted $G_{\faults}$.

\medskip
A key technical ingredient in the MWPM decoder is the {\em distance graph} of a subset of vertices $\bar \sigma \subset V(G_{\faults})$ of the decoding graph. The distance graph $K_{\bar \sigma}$ is the graph whose vertices correspond to the elements of $\bar \sigma$.
Two vertices of $K_{\bar \sigma}$ are connected by an edge iff they live in the same connected component of the decoding graph $G_{\faults}$.
Moreover, the weight of this edge is given by the weighted distance between these vertices in $G_{\faults}$.
The MWPM decoder takes as an input a syndrome and returns a most likely fault configuration by computing a minimum-weight perfect matching in the distance graph. This can be done in polynomial time thanks to Edmond's algorithm~\cite{edmonds1965maximum, edmonds1965paths}.

\medskip
With these assumptions, the MWPM decoder (Algorithm~\ref{algo:mwpm_decoder}) computes a most likely fault configuration. The Union-Find (UF) decoder~\cite{delfosse2021almost} can be built from the same decoding graph (without using the distance graph). It provides a good approximation of the MWPM decoder with a more favorable complexity.

\medskip
Let $\faults$ be a set of faults that satisfies assumptions~\ref{assumption:edges_like_fault} and \ref{assumption:linearity}. The MWPM decoder and the UF decoder associated with $\faults$ are denoted $\MWPM_{\faults}$ and $\UF_{\faults}$. Given a syndrome $\sigma \subset \checks$, the fault configuration returned by the decoder is denoted $\MWPM_{\faults}(\sigma)$ or $\UF_{\faults}(\sigma)$.

\subsection{Examples}

A classical memory encoded with the repetition code which suffers from independent bit-flips is an example which satisfies these two assumptions.
A bit $x=0$ or 1 is encoded in a bit string $(x, x, \dots, x)$ with $n$ repetitions. It comes with $n-1$ checks that compute the parities of two consecutive bits: $x_i + x_{i+1} \pmod 2$ for $i=0, \dots, n-2$.
By definition, checks are linear and a single bit-flip triggers either one or two checks.

\medskip
The surface code~\cite{dennis2002topological} with perfect measurements and $X$ faults or $Z$ faults is another example.
Each plaquette measurement defines a check. The plaquette outcomes are linear and each $X$ fault triggers the two incident $Z$ plaquettes (only one for boundary qubits).  Similarly, each $Z$ fault triggers two incident $X$ plaquettes.

\medskip
Phenomenological measurement noise in the surface code~\cite{dennis2002topological} also satisfies assumptions \ref{assumption:edges_like_fault} and \ref{assumption:linearity}.
When measurements are noisy, we repeat plaquette measurements to correct their outcomes. Assume that we run $T$ consecutive rounds of measurement and that each round of measurement is followed by a round of independent $X$ faults on the code qubits.
A check is not anymore the outcome of a single plaquette. Instead, there is a check for each plaquette $i$ and each time step $t = 0, \dots T-1$. The value of the check $(i, t)$ is defined to be 1 iff the outcome of plaquette $i$ changes between time step $t-1$ and $t$. To define the check value for $t=0$, we assume that the outcomes at time step $t=-1$ are all 0.
An $X$ fault occurring after time step $t$ triggers the checks corresponding to the (at most two) incident plaquettes at time step $t+1$.
The flip of the outcome of plaquette $i$ at time step $t$ triggers the checks $(i, t)$ and $(i, t+1)$. Such a flip triggers only one check when $t=0$ or $T-1$.

\medskip
The circuit noise model with $X$ faults for the surface code with standard plaquette measurement circuits based on CNOT gates~\cite{fowler2012surface} or joint measurements~\cite{chao2020optimization} also satisfies assumptions \ref{assumption:edges_like_fault} and \ref{assumption:linearity}.

\medskip
For the standard syndrome extraction circuits, the only type of fault that is problematic for MWPM decoding of surface codes is $Y$ faults because they trigger either three or four checks. However, each $Y$ fault naturally decomposes as a product of an $X$ fault and a $Z$ fault. One can correct all Pauli faults and outcome flips with the surface codes by correcting independently $X$ faults and $Z$ faults. This leads to a MWPM decoder that achieves the full distance of the surface code. One can improve this strategy using the correlations between $X$ and $Z$~\cite{fowler2013optimal, delfosse2014decoding}.

\section{Splitting noise models}
\label{sec:splitting}

Floquet codes are more difficult to decode because some faults induce weight four syndromes.
Consider, for instance, Floquet codes defined on a toric lattice~\cite{hastings2021dynamically}.
There are four types of faults: $X$ faults, $Y$ faults, $Z$ faults and measurement outcome flips.
The three types of single qubit Pauli faults trigger two checks, but measurement flips trigger four checks.
In the case of surface codes, there is a natural split of $Y$ faults as $Y=XZ$ into a pair of faults that satisfy assumption~\ref{assumption:edges_like_fault}. 
The splitting of measurement flip is less obvious for Floquet codes~\footnote{One can split a measurement fault by considering the spacetime picture as follows. The flip of the outcome of a two-qubit measurement $X_iX_j$ is equivalent to a Pauli fault $Z_i$ right before the measurement and a Pauli fault $Z_i$ right after the measurement.}.
Here, we describe a splitting strategy that applies to both surface codes and Floquet codes. 
Combined with the MWPM decoder or the UF decoder this leads to an efficient decoder that reaches the largest achievable distance for standard surface codes and Floquet surface codes. 

\subsection{Primitive faults}

\medskip
Define a $w$-fault to be a fault that triggers $w$ checks.
Clearly, $0$-faults are undetectable and therefore not correctable. We assume that none of the faults $f_i$ defining the noise model is a $0$-fault.

\medskip
Given a noise model with independent faults $\faults = \{f_1, \dots, f_m\}$, a fault $f_i$ is said to be {\em primitive} if it is a 1-fault or if it is a 2-fault and if its syndrome is not the sum of two 1-fault syndromes.
The set of primitive faults is denoted $\faults' \subset \faults$.

\medskip
Primitive faults satisfy the two assumptions required for the standard MWPM decoder. We can therefore build a decoding graph from the set of primitive faults and define a MWPM decoder or a UF decoder using this graph.

\medskip
The set of primitive faults does not contain all the faults of $\faults$ which satisfy assumptions \ref{assumption:edges_like_fault}. For surface codes, a $Y$ fault at the corner of the lattice is a 2-fault but is not a primitive fault because it is a product of an $X$ fault and a $Z$ fault which are 1-faults.
We do not include this $Y$ fault in the set of primitive faults because it would reduce the effective distance of the decoder by creating a shortcut in the decoding graph.

\subsection{Decoder-based splitting}

The graph induced by primitive faults is used in combination with the standard MWPM decoder to split non-primitive faults into 1-faults and 2-faults as explained in Algorithm~\ref{algo:fault_splitting}.
The whole procedure is represented in Figure~\ref{fig:splitting_diagram}.
A non-primitive fault $f$ is decomposed by calling the MWPM decoder $\MWPM_{\faults'}$ associated with primitive faults.
This produces a set of fault configurations $D_f = \{\faultset_1, \dots, \faultset_s\}$ such that each fault $\faultset_i$ is either a 1-fault or a 2-fault. 
Moreover, the syndrome of the sum $\faultset_1 + \dots + \faultset_s$ is the syndrome of $f$.
This decomposition allows us to split non-primitive faults into 1-faults and 2-faults that can be added to the set of primitive faults. 
To speed up the fault decomposition, we could replace $\MWPM_{\faults'}$ by the Union-Find decoder $\UF_{\faults'}$ in Algorithm~\ref{algo:fault_splitting}.

\begin{algorithm}
\DontPrintSemicolon
\SetKwInOut{Input}{input}\SetKwInOut{Output}{output}
\Input{A fault configuration $\faultset$ with syndrome $\sigma$. The MWPM decoder $\MWPM_{\faults'}$ based on primitive faults.}
\Output{A set of fault configurations $D_\varphi = \{\varphi_1, \dots, \varphi_s\}$ such that
(i) Each fault configuration $\varphi_i$ is either a 1-fault or a 2-fault and
(ii) The syndrome of $\varphi_1  + \dots + \varphi_s$ is $\sigma$.}

\BlankLine
	Compute $\tilde f = \MWPM_{\faults'}(\sigma)$. \;
	Compute $\bar \sigma$ as in Algorithm~\ref{algo:mwpm_decoder}. \;
	Partition $\tilde f$ into disjoint paths $\pi_1, \dots, \pi_s$ whose endpoints are the vertices of $\bar \sigma$. \;
	Initialize $D_\varphi = \{\}$. \;
	\For {each path $\pi$}
	{
		Let $e_{i_1}, e_{i_2}, \dots, e_{i_t}$ be the edges of the path $\pi$. \;
		Add $\faultset_i$ to $D_\varphi$ where $\faultset_i = f_{i_1} + f_{i_2} + \dots + f_{i_t}$. \;
	}
	Return $D_\varphi$. \;
	\caption{Splitting of a fault configuration.}
	\label{algo:fault_splitting}
\end{algorithm}

\medskip
Given a noise model with independent faults $\faults$, we construct a {\em split noise model} with independent faults $\faults''$ as explained in Algorithm~\ref{algo:noise_splitting}.
First, we add all the primitive faults of $\faults$ to $\faults''$.
Then, we loop over the non-primitive faults and for each non-primitive fault $f$, we compute the decomposition $D_f$ of $f$ using Algorithm~\ref{algo:fault_splitting} and we add each fault of $D_f$ to $\faults''$ with corresponding probability $p$ (the initial probability of $f$).
The resulting set of faults $\faults''$ satisfies assumptions \ref{assumption:edges_like_fault} and \ref{assumption:linearity}.
We can therefore define a MWPM decoder or a UF decoder based on the split noise model $\faults''$.
One can interpret $\faults''$ as an approximation of the noise model $\faults$ by graph-like noise model.

\begin{algorithm}
\DontPrintSemicolon
\SetKwInOut{Input}{input}\SetKwInOut{Output}{output}
\Input{A noise model $\faults$.}
\Output{A graph-like noise model $\faults''$.}

\BlankLine
	Compute the set of primitive faults $\faults'$ of $\faults$. \;
	Construct the MWPM decoder $\MWPM_{\faults'}$. \;
	Initialize the noise model $\faults'' = \faults'$. \;
	\For {each non-primitive fault $f$ with probability $p_f$}
	{
		Compute the decomposition $D_f$ using Algorithm~\ref{algo:fault_splitting}. \;
		\For {each fault $\faultset$ in $D_f$}
		{
			Add $\faultset$ with corresponding probability $p_{\faultset} = p_f$ to the noise model $\faults''$. \;
		}
	}
	Return the noise model $\faults''$. \;
	\caption{Decoder-based splitting.}
	\label{algo:noise_splitting}
\end{algorithm}

\medskip
We used this strategy to decode the Floquet surface codes in~\cite{paetznick2022performance} and observed numerically that it achieves the maximum distance achievable for the hexagon and square-octagon lattices.
This idea also leads to decoders that achieve the full code distance of the surface codes with different noise models (perfect measurement, phenomenological, circuit noise) and different syndrome extraction circuit (CNOT-based~\cite{fowler2012surface}, measurement-based~\cite{chao2020optimization}).
The strength of this approach is its flexibility which makes it a convenient tool to quickly explore the performance of the new variants of topological codes, new boundary conditions or new circuits without the need to design a new decoder.

\medskip
This idea only applies to codes and noise models with a specific structure. 
For example, it does not work with color codes on a torus with perfect measurements because, in this case, the set of primitive faults is empty. 
This is because each color code fault triggers exactly three checks.
It may also happen that some non-primitive faults cannot be decomposed into primitive faults by Algorithm~\ref{algo:fault_splitting} because some checks triggered by this fault are not triggered by any of the primitive faults.

\subsection{Recursive splitting}

\begin{algorithm}
\DontPrintSemicolon
\SetKwInOut{Input}{input}\SetKwInOut{Output}{output}
\Input{A noise model $\faults$.}
\Output{A graph-like noise model $\faults''$.}

\BlankLine
	Initialize $\faults'' = \{\}$. \;
	\While {$\faults$ is not empty and the following loop is not trivial}
	{
		\For {each $w=1, 2, \dots, $}
		{
			\For {each $w$-fault $f$ of $\faults$}
			{
				\If {$f$ is a 1-fault}
				{
					Remove $f$ from $\faults$. \;
					Add $f$ to $\faults''$. \;
				}
				\If {$f$ is a 2-fault and $\sigma(f)$ is not the sum of the syndromes of two 1-faults of $\faults''$}
				{
					Remove $f$ from $\faults$. \;
					Add $f$ to $\faults''$. \;
				}
				\If {there exists a fault $g \in \faults''$ such that $\sigma(g) \subset \sigma(f)$}
				{
					Define the fault $h$ with $\sigma(h) = \sigma(f) \backslash \sigma(g)$ and with probability $\Prob(h) = \Prob(f)$. \;
					Remove $f$ from $\faults$. \;
					Add $h$ to $\faults$. \;
				}
			}
		}
	}
	Return $\faults''$. \;
	\caption{Recursive Splitting.}
	\label{algo:alternative_split}
\end{algorithm}

Here, we discuss an alternative splitting strategy decribed in Algorithm~\ref{algo:alternative_split}.
Its main advantage over Algorithm~\ref{algo:noise_splitting} is that it is simpler and it does not need a decoder.
Neither strategy is strictly better than the other in the sense that there exist faults that can be split by one of the algorithms and not by the other.
These two splitting algorithms can be combined to extend the range of application of the MWPM decoder.

\medskip
The basic idea of Algorithm~\ref{algo:alternative_split} is to split a fault $f$ by removing the primitive parts of $f$ until nothing remains.
In general, it provides the same decomposition of $Y$ faults in the surface codes and outcome flips in Floquet codes as the previous strategy. 
However, Algorithm~\ref{algo:fault_splitting} fails to decompose a $3$-fault whose syndrome is of the form $\{a, b, c\}$ where $a$ and $b$ appear in the syndrome of primitive faults but $c$ does not.
On the contrary Algorithm~\ref{algo:alternative_split} succeeds to split this fault.
A limitation of Algorithm~\ref{algo:alternative_split} is that it cannot always split faults that are the product of paths where each path contains at least two primitive faults. Algorithm~\ref{algo:fault_splitting} works well in this case.

\medskip
Splitting a noise model may produce a split model which includes multiple copies of the same fault. We can combine these copies of the same fault as discussed in Section~\ref{subsec:faults}.

\medskip
One could consider different variant of Algorithm~\ref{algo:alternative_split}. For example, instead of a while loop, we could use a heap to prioritize the faults with minimum syndrome weight and update the position of a fault after the removal of a component $g$ of a fault $f$.
We wrote the pseudo-code of Algorithm~\ref{algo:alternative_split} with multiple nested loops to make it easy to read and to understand.
A more efficient implementation can be obtained by exploiting the exact structure of the set of faults. In particular, for a noise model with faults that triggers a small number of checks, we could use the Tanner graph~\cite{tanner1981recursive} of the noise model to rapidly check the conditions in line 8 and 12 of Algorithm~\ref{algo:alternative_split}.

\medskip
Finally, it seems natural to combine our two splitting methods. We could first generate primitive faults using the strategy of Algorithm~\ref{algo:alternative_split} and then split the remaining non-primitive faults using Algorithm~\ref{algo:noise_splitting}.
We could use Algorithm~\ref{algo:noise_splitting} first before Algorithm~\ref{algo:alternative_split}.

\section{Conclusion}

Decoding is hard~\cite{berlekamp1978inherent, iyer2015hardness} and we do not expect the decoding problem for a general code to be efficiently solvable.
For graph-like noise models, faults can be interpreted as edges in a graph and the decoding problem can be solved efficiently by reducing it to a matching problem in a graph.
This is the case of surface codes and repetition codes with the MWPM decoder or the UF decoder.
We proposed two different heuristic strategies allowing us to apply these decoders to hypergraphs by splitting hyperedges into edges and we observe numerically that these decoders achieve the maximum achievable distance for the hypergraph corresponding to the decoding problem of some Floquet codes.
Our splitting decoder could be relevant to explore numerically the performance of other recent variants of Floquet codes~\cite{aasen2022adiabatic, davydova2023floquet, kesselring2022anyon, bombin2023unifying, townsend2023floquetifying, dua2023engineering, zhang2022x, ellison2023floquet, davydova2023quantum}.

Not all LDPC codes admit a splitting decoder. Consider an expander graph $G$ and define a classical code by placing bits on the vertices of the graph and checks on the edges. The check supported on a edge $\{u, v\}$ is the sum of the two bits supported on $u$ and $v$. Any error pattern corresponds to some set $S$ of flipped vertices, and the violated checks are the boundary of the set of flipped vertices: the violated checks go from vertices in $S$ to those not in $S$.
If the graph is a good enough expander, no set $S$ has only one or two edges in its boundary, proving that there is no splitting for this code.
 
In future work, one may try to identify a set of sufficient conditions which guarantee that the splitting decoder achieves the full code distance of a given LDPC code.
We may also try to bound the gap between the code distance and the distance achieved by the splitting decoder as a function of the Tanner graph of the code.
If this gap is sufficiently small, the decoder can still achieve a good performance in practice, even if it does not reach the full code distance.

\section*{Acknowledgment}

We would like to thank Dave Aasen, Michael Beverland, Vadym Kliuchnikov, Marcus Silva, Shilin Huang for their comments on a preliminary version of this work.

%

\begin{thebibliography}{10}

\bibitem{aasen2022adiabatic}
David Aasen, Zhenghan Wang, and Matthew~B Hastings.
\newblock Adiabatic paths of hamiltonians, symmetries of topological order, and
  automorphism codes.
\newblock {\em Physical Review B}, 106(8):085122, 2022.

\bibitem{berlekamp1978inherent}
Elwyn Berlekamp, Robert McEliece, and Henk Van~Tilborg.
\newblock On the inherent intractability of certain coding problems (corresp.).
\newblock {\em IEEE Transactions on Information Theory}, 24(3):384--386, 1978.

\bibitem{bombin2023unifying}
Hector Bombin, Daniel Litinski, Naomi Nickerson, Fernando Pastawski, and Sam
  Roberts.
\newblock Unifying flavors of fault tolerance with the {ZX} calculus.
\newblock {\em arXiv preprint arXiv:2303.08829}, 2023.

\bibitem{breuckmann2021quantum}
Nikolas~P Breuckmann and Jens~Niklas Eberhardt.
\newblock Quantum low-density parity-check codes.
\newblock {\em PRX Quantum}, 2(4):040101, 2021.

\bibitem{chao2020optimization}
Rui Chao, Michael~E Beverland, Nicolas Delfosse, and Jeongwan Haah.
\newblock Optimization of the surface code design for {Majorana}-based qubits.
\newblock {\em Quantum}, 4:352, 2020.

\bibitem{davydova2023floquet}
Margarita Davydova, Nathanan Tantivasadakarn, and Shankar Balasubramanian.
\newblock {Floquet} codes without parent subsystem codes.
\newblock {\em PRX Quantum}, 4(2):020341, 2023.

\bibitem{davydova2023quantum}
Margarita Davydova, Nathanan Tantivasadakarn, Shankar Balasubramanian, and
  David Aasen.
\newblock Quantum computation from dynamic automorphism codes.
\newblock {\em arXiv preprint arXiv:2307.10353}, 2023.

\bibitem{delfosse2022toward}
Nicolas Delfosse, Vivien Londe, and Michael~E Beverland.
\newblock Toward a union-find decoder for quantum {LDPC} codes.
\newblock {\em IEEE Transactions on Information Theory}, 68(5):3187--3199,
  2022.

\bibitem{delfosse2021almost}
Nicolas Delfosse and Naomi~H Nickerson.
\newblock Almost-linear time decoding algorithm for topological codes.
\newblock {\em Quantum}, 5:595, 2021.

\bibitem{delfosse2023spacetime}
Nicolas Delfosse and Adam Paetznick.
\newblock Spacetime codes of {Clifford} circuits.
\newblock {\em arXiv preprint arXiv:2304.05943}, 2023.

\bibitem{delfosse2014decoding}
Nicolas Delfosse and Jean-Pierre Tillich.
\newblock A decoding algorithm for {CSS} codes using the {X/Z} correlations.
\newblock In {\em 2014 IEEE International Symposium on Information Theory},
  pages 1071--1075. IEEE, 2014.

\bibitem{dennis2002topological}
Eric Dennis, Alexei Kitaev, Andrew Landahl, and John Preskill.
\newblock Topological quantum memory.
\newblock {\em Journal of Mathematical Physics}, 43(9):4452--4505, 2002.

\bibitem{du2022stabilizer}
Julien Du~Crest, Mehdi Mhalla, and Valentin Savin.
\newblock Stabilizer inactivation for message-passing decoding of quantum
  {LDPC} codes.
\newblock In {\em 2022 IEEE Information Theory Workshop (ITW)}, pages 488--493.
  IEEE, 2022.

\bibitem{dua2023engineering}
Arpit Dua, Nathanan Tantivasadakarn, Joseph Sullivan, and Tyler~D Ellison.
\newblock Engineering {Floquet} codes by rewinding.
\newblock {\em arXiv preprint arXiv:2307.13668}, 2023.

\bibitem{edmonds1965maximum}
Jack Edmonds.
\newblock Maximum matching and a polyhedron with 0, 1-vertices.
\newblock {\em Journal of research of the National Bureau of Standards B},
  69(125-130):55--56, 1965.

\bibitem{edmonds1965paths}
Jack Edmonds.
\newblock Paths, trees, and flowers.
\newblock {\em Canadian Journal of mathematics}, 17:449--467, 1965.

\bibitem{ellison2023floquet}
Tyler~D Ellison, Joseph Sullivan, and Arpit Dua.
\newblock {Floquet} codes with a twist.
\newblock {\em arXiv preprint arXiv:2306.08027}, 2023.

\bibitem{fowler2013optimal}
Austin~G Fowler.
\newblock Optimal complexity correction of correlated errors in the surface
  code.
\newblock {\em arXiv preprint arXiv:1310.0863}, 2013.

\bibitem{fowler2012surface}
Austin~G Fowler, Matteo Mariantoni, John~M Martinis, and Andrew~N Cleland.
\newblock Surface codes: Towards practical large-scale quantum computation.
\newblock {\em Physical Review A}, 86(3):032324, 2012.

\bibitem{fowler2012towards}
Austin~G Fowler, Adam~C Whiteside, and Lloyd~CL Hollenberg.
\newblock Towards practical classical processing for the surface code.
\newblock {\em Physical review letters}, 108(18):180501, 2012.

\bibitem{gidney2021stim}
Craig Gidney.
\newblock Stim: a fast stabilizer circuit simulator.
\newblock {\em Quantum}, 5:497, 2021.

\bibitem{gidney2022benchmarking}
Craig Gidney, Michael Newman, and Matt McEwen.
\newblock Benchmarking the planar honeycomb code.
\newblock {\em Quantum}, 6:813, 2022.

\bibitem{grospellier2021combining}
Antoine Grospellier, Lucien Grou{\`e}s, Anirudh Krishna, and Anthony Leverrier.
\newblock Combining hard and soft decoders for hypergraph product codes.
\newblock {\em Quantum}, 5:432, 2021.

\bibitem{gu2023efficient}
Shouzhen Gu, Christopher~A Pattison, and Eugene Tang.
\newblock An efficient decoder for a linear distance quantum {LDPC} code.
\newblock In {\em Proceedings of the 55th Annual ACM Symposium on Theory of
  Computing}, pages 919--932, 2023.

\bibitem{hastings2021dynamically}
Matthew~B Hastings and Jeongwan Haah.
\newblock Dynamically generated logical qubits.
\newblock {\em Quantum}, 5:564, 2021.

\bibitem{hastings2021fiber}
Matthew~B Hastings, Jeongwan Haah, and Ryan O'Donnell.
\newblock Fiber bundle codes: breaking the n 1/2 polylog (n) barrier for
  quantum {LDPC} codes.
\newblock In {\em Proceedings of the 53rd Annual ACM SIGACT Symposium on Theory
  of Computing}, pages 1276--1288, 2021.

\bibitem{higgott2023constructions}
Oscar Higgott and Nikolas~P Breuckmann.
\newblock Constructions and performance of hyperbolic and semi-hyperbolic
  floquet codes.
\newblock {\em arXiv preprint arXiv:2308.03750}, 2023.

\bibitem{higgott2023sparse}
Oscar Higgott and Craig Gidney.
\newblock Sparse blossom: correcting a million errors per core second with
  minimum-weight matching.
\newblock {\em arXiv preprint arXiv:2303.15933}, 2023.

\bibitem{iyer2015hardness}
Pavithran Iyer and David Poulin.
\newblock Hardness of decoding quantum stabilizer codes.
\newblock {\em IEEE Transactions on Information Theory}, 61(9):5209--5223,
  2015.

\bibitem{kesselring2022anyon}
Markus~S Kesselring, Julio C~Magdalena de~la Fuente, Felix Thomsen, Jens
  Eisert, Stephen~D Bartlett, and Benjamin~J Brown.
\newblock Anyon condensation and the color code.
\newblock {\em arXiv preprint arXiv:2212.00042}, 2022.

\bibitem{kovalev2012improved}
Alexey~A Kovalev and Leonid~P Pryadko.
\newblock Improved quantum hypergraph-product {LDPC} codes.
\newblock In {\em 2012 IEEE International Symposium on Information Theory
  Proceedings}, pages 348--352. IEEE, 2012.

\bibitem{leverrier2022quantum}
Anthony Leverrier and Gilles Z{\'e}mor.
\newblock Quantum tanner codes.
\newblock In {\em 2022 IEEE 63rd Annual Symposium on Foundations of Computer
  Science (FOCS)}, pages 872--883. IEEE, 2022.

\bibitem{leverrier2023efficient}
Anthony Leverrier and Gilles Z{\'e}mor.
\newblock Efficient decoding up to a constant fraction of the code length for
  asymptotically good quantum codes.
\newblock In {\em Proceedings of the 2023 Annual ACM-SIAM Symposium on Discrete
  Algorithms (SODA)}, pages 1216--1244. SIAM, 2023.

\bibitem{lin2022good}
Ting-Chun Lin and Min-Hsiu Hsieh.
\newblock Good quantum {LDPC} codes with linear time decoder from lossless
  expanders.
\newblock {\em arXiv preprint arXiv:2203.03581}, 2022.

\bibitem{mackay2003information}
David~JC MacKay.
\newblock {\em Information theory, inference and learning algorithms}.
\newblock Cambridge university press, 2003.

\bibitem{paetznick2022performance}
Adam Paetznick, Christina Knapp, Nicolas Delfosse, Bela Bauer, Jeongwan Haah,
  Matthew~B Hastings, and Marcus~P da~Silva.
\newblock Performance of planar {Floquet} codes with {Majorana}-based qubits.
\newblock {\em arXiv preprint arXiv:2202.11829}, 2022.

\bibitem{panteleev2021degenerate}
Pavel Panteleev and Gleb Kalachev.
\newblock Degenerate quantum {LDPC} codes with good finite length performance.
\newblock {\em Quantum}, 5:585, 2021.

\bibitem{panteleev2022asymptotically}
Pavel Panteleev and Gleb Kalachev.
\newblock Asymptotically good quantum and locally testable classical {LDPC}
  codes.
\newblock In {\em Proceedings of the 54th Annual ACM SIGACT Symposium on Theory
  of Computing}, pages 375--388, 2022.

\bibitem{poulin2008iterative}
David Poulin and Yeojin Chung.
\newblock On the iterative decoding of sparse quantum codes.
\newblock {\em arXiv preprint arXiv:0801.1241}, 2008.

\bibitem{raussendorf2007fault}
Robert Raussendorf and Jim Harrington.
\newblock Fault-tolerant quantum computation with high threshold in two
  dimensions.
\newblock {\em Physical review letters}, 98(19):190504, 2007.

\bibitem{richardson2008modern}
Tom Richardson and Ruediger Urbanke.
\newblock {\em Modern coding theory}.
\newblock Cambridge university press, 2008.

\bibitem{roffe2020decoding}
Joschka Roffe, David~R White, Simon Burton, and Earl Campbell.
\newblock Decoding across the quantum low-density parity-check code landscape.
\newblock {\em Physical Review Research}, 2(4):043423, 2020.

\bibitem{tanner1981recursive}
R~Tanner.
\newblock A recursive approach to low complexity codes.
\newblock {\em IEEE Transactions on information theory}, 27(5):533--547, 1981.

\bibitem{tillich2013quantum}
Jean-Pierre Tillich and Gilles Z{\'e}mor.
\newblock Quantum {LDPC} codes with positive rate and minimum distance
  proportional to the square root of the blocklength.
\newblock {\em IEEE Transactions on Information Theory}, 60(2):1193--1202,
  2013.

\bibitem{townsend2023floquetifying}
Alex Townsend-Teague, Julio~Magdalena de~la Fuente, and Markus Kesselring.
\newblock Floquetifying the colour code.
\newblock {\em arXiv preprint arXiv:2307.11136}, 2023.

\bibitem{tremblay2022constant}
Maxime~A Tremblay, Nicolas Delfosse, and Michael~E Beverland.
\newblock Constant-overhead quantum error correction with thin planar
  connectivity.
\newblock {\em Physical Review Letters}, 129(5):050504, 2022.

\bibitem{zhang2022x}
Zhehao Zhang, David Aasen, and Sagar Vijay.
\newblock The x-cube {Floquet} code.
\newblock {\em arXiv preprint arXiv:2211.05784}, 2022.

\end{thebibliography}

\end{document}